\RequirePackage{fix-cm}
\documentclass[smallextended]{svjour3}       
\smartqed  
\usepackage{graphicx}
\usepackage[figuresright]{rotating}
\usepackage{subfig}
\usepackage{amsmath,amssymb}
\usepackage{algorithm,algorithmic}

\usepackage{cite}

\usepackage{color}

\newcommand{\be}{\begin{equation}}
\newcommand{\ee}{\end{equation}}
\newcommand{\beqy}{\begin{eqnarray}}
\newcommand{\eeqy}{\end{eqnarray}}
\newcommand{\beqynn}{\begin{eqnarray*}}
\newcommand{\eeqynn}{\end{eqnarray*}}

\newcommand{\ba}{\begin{array}}
\newcommand{\ea}{\end{array}}
\newcommand{\bmx}{\begin{bmatrix}}
\newcommand{\emx}{\end{bmatrix}}
\newcommand{\bsmx}{\left[\begin{smallmatrix}}
\newcommand{\esmx}{\end{smallmatrix}\right]}
\newcommand{\bmxc}[1]{\left[\begin{array}{@{}#1@{}}}
\newcommand{\emxc}{\end{array}\right]}

\newcommand{\bt}[1]{\begin{tabular}{#1}}
\newcommand{\et}{\end{tabular}}

\newcommand{\bc}{\begin{center}}
\newcommand{\ec}{\end{center}}
\newcommand{\ben}{\begin{enumerate}}
\newcommand{\een}{\end{enumerate}}
\newcommand{\bi}{\begin{itemize}}
\newcommand{\ei}{\end{itemize}}

\newcommand{\A}{\boldsymbol{A}}
\newcommand{\B}{\boldsymbol{B}}

\newcommand{\I}{\boldsymbol{I}}

\renewcommand{\P}{\boldsymbol{P}}

\newcommand{\X}{{\boldsymbol{X}}}

\renewcommand{\b}{\boldsymbol{b}}

\renewcommand{\d}{\boldsymbol{d}}

\renewcommand{\v}{{\boldsymbol{v}}}

\newcommand{\x}{{\boldsymbol{x}}}
\newcommand{\y}{{\boldsymbol{y}}}
\newcommand{\z}{{\boldsymbol{z}}}

\newcommand{\0}{{\boldsymbol{0}}}

\begin{document}

\title{A Novel Approach for Ellipsoidal Outer-Approximation of the Intersection Region of Ellipses in the Plane
\thanks{Funding for this work was provided in parts by research grants from the Natural Sciences and Engineering Research Council of Canada}
}
\subtitle{}


\author{Siamak Yousefi \and
        Xiao-Wen Chang \and
        Henk Wymeersch \and
        Benoit Champagne \and
        Godfried Toussaint 
}


\institute{S. Yousefi \at
              Department of Electrical and Computer Engineering, McGill University, Montreal, QC, Canada \\
              \email{siamak.yousefi@mail.mcgill.ca}           
           \and
           X.-W. Chang \at
              School of Computer Science, McGill University, Montreal, QC, Canada\\       
              \email{chang@cs.mcgill.ca}
            \and
            H. Wymeersch \at
            Department of Signals and Systems, Chalmers University of Technology, Gothenburg, Sweden\\ 
            \email{henkw@chalmers.se}          
            \and
            B. Champagne \at
            Department of Electrical and Computer Engineering, McGill University, Montreal, QC, Canada \\
            \email{benoit.champagne@mcgill.ca} 
            \and
            G. Toussaint \at
            Computer Science Program, New York University Abu Dhabi, United Arab Emirates \\
            \email{gt42@nyu.edu}
}

\date{Received: date / Accepted: date}

\maketitle

\begin{abstract}
In this paper, a novel technique for tight outer-approximation of the intersection region of a finite number of ellipses in 2-dimensional (2D) space is proposed.
First, the vertices of a tight polygon that contains the convex intersection of the ellipses are found in an efficient manner.
To do so, the intersection points of the ellipses that fall on the boundary of the intersection region are determined, and a set of points is generated on the elliptic arcs connecting every two neighbouring intersection points.
By finding the tangent lines to the ellipses at the extended set of points, a set of half-planes is obtained, whose intersection forms a polygon.
To find the polygon more efficiently, the points are given an order and the intersection of the half-planes corresponding to every two neighbouring points is calculated.
If the polygon is convex and bounded, these calculated points together with the initially obtained intersection points will form its vertices. 
If the polygon is non-convex  or unbounded, we can detect this situation and then generate additional discrete points only on the elliptical arc segment causing the issue, and restart the algorithm to obtain a bounded and convex polygon.
Finally, the smallest area ellipse that contains the vertices of the polygon is obtained by solving a convex optimization problem.
Through numerical experiments, it is illustrated that the proposed technique returns a tighter outer-approximation of the intersection of multiple ellipses, compared to conventional techniques, with only slightly higher computational cost.
\end{abstract}

\section{Introduction}
\label{intro}
In many areas of science and engineering, such as computational geometry, image processing, control systems, parameter estimation, and wireless communications, a complex convex set needs to be represented by a simpler geometric shape containing it, e.g., see in \cite{Enclosing_1, Enclosing_2, Enclosing_3, Enclosing_4, Enclosing_5, Enclosing_6, Enclosing_7, Book_Ellipsoidal_Calculus, Boyd_LMI_Book} and the references therein.
Among possible different shapes, ellipsoids are often considered since they can be easily described in terms of vectors and matrices. 
Ellipsoids often provide tight outer-approximations of the underlying convex sets, and are invariant under affine transformations \cite{Book_Ellipsoidal_Calculus}.
Therefore, ellipsoidal calculus has gained significant attention in many different fields due to its importance and usefulness.
For instance, in control theory, ellipsoidal bounds are used to describe the uncertainty of sets associated with state space models \cite{State_Bounding_Ellipsoid}.
In sensor network localization, ellipsoids are employed to provide constraints on the unknown positions of sensors \cite{Gholami_Bounding}, \cite{Zhang_Ellipse_Outer}.

According to \cite[p.44]{Boyd_LMI_Book}, verifying that an ellipsoid covers the intersection of a given number of ellipsoids is NP-complete \textbf{in the space dimension}, so the problem cannot be expressed as a linear matrix inequality (LMI).
As a result, finding the minimum volume ellipsoid which contains the intersection of multiple ellipsoids may not be recast as a convex optimization problem \textbf{in general}.
To the authors' knowledge, \textbf{in general}, there exists no practically efficient computer code to find the optimal solution to this problem, and all the tested proposed techniques offer sub-optimal solutions. However, it is possible that the framework of LP-type problems might be suitable to solve this problem optimally in theory \cite{Subexponential_bound}.
Earlier work on this topic focused on the special case of two ellipsoids due to its simplicity.
For instance in \cite{2_Ellipse_Halfspace}, one of the ellipsoids is approximated by a half-space, and then the problem of finding the tightest ellipsoid containing the intersection of a half-space and an ellipsoid can be solved optimally.
It is mentioned in \cite{2_Ellipse_Optimal} that the optimal ellipsoid can be expressed as a linear convex combination of the two ellipsoids \textbf{when the two have the same center.}
The optimization problem for searching within the convex combinations of two ellipsoids is investigated in \cite{State_Bounding_Ellipsoid} and further analysis and theoretical results are derived in \cite{Bounding_Ellipsoid_2}.
For a larger number of ellipsoids, standard convex optimization techniques can be used in order to efficiently find an outer-approximation, as summarized in \cite{Boyd_LMI_Book}.
For instance, one well known technique is to first obtain the largest ellipsoid enclosed in the intersection of the given ellipsoids, which can be done optimally by solving a convex optimization problem \cite{Boyd}.
Then, as shown by John and Lowner (see \cite{John_Lowner} and the references therein), upon scaling the calculated ellipsoid by the dimension of the space, the resulting ellipsoid is guaranteed to contain the intersection region.
%
These techniques are approximate and suboptimal, although bounding ellipses can be obtained in polynomial time using these convex optimization techniques.
We will show this fact through simulations.

It is clear that the problem is quite challenging for arbitrary dimensions, but for 2-dimensional (2-D) space, low-cost geometrical techniques can be developed.
Recently, in the context of sensor network localization under non-line-of-sight (NLOS) propagation, the authors in \cite{Siamak_Letter_BP} proposed a 2-stage method for tight outer-approximation of the intersection of multiple ellipses in 2-D space, and then plugged-in their technique to the distributed bounding algorithm developed in \cite{Gholami_Bounding}.
In this method, first a polygon is obtained by generating discrete points on the boundary of each ellipse, rejecting those that fall outside the feasible region, and then intersecting the half planes tangent to the ellipses at those remaining points. 
Subsequently, the tightest ellipse that contains the vertices of the resulting polygon is obtained by solving a convex optimization problem \cite{Min_Vol_Ellipse_Points}.
This method can generally outperform existing approaches in the literature (see \cite{Boyd_LMI_Book}) in 2-D, if enough discrete points are generated on each ellipse; however, it is always possible that the polygon becomes unbounded or non-convex.
Although this problem can be avoided by generating a large number of discrete points on each ellipse, this will render the method inefficient when the number of ellipses is large.

In this work, we extend the work of \cite{Siamak_Letter_BP} for outer-approximation of the intersection of several ellipses in 2-D space in the following ways, such that a tight bounded and convex polygon can be formed, and the drawbacks mentioned above can be avoided: 

\begin{itemize}

\item For every elliptic arc forming the boundary of the intersection of ellipses, we find the two end points, which are the intersection points of two or more ellipses, and generate a desired number of points on that arc.
Compared to \cite{Siamak_Letter_BP}, in this way, we avoid generating unnecessary discrete points, and thus improve the efficiency.

\item The tangent lines corresponding to every two neighbouring points intersect, leading to a new set of points.
These points along with the intersection points of the ellipses obtained earlier are used as the vertices of the possibly bounded and convex polygon.

\item The boundedness and convexity of the polygon are verified and if it is unbounded or non-convex, the number of discrete points on the elliptic arc which causes the problem is increased.
Therefore, the new method will eventually find a bounded convex polygon containing the intersection of ellipses.

\end{itemize}
Finally, the tightest ellipse containing the vertices of the generated polygon is obtained by solving a convex optimization problem.
Through numerical evaluations we observe that the proposed method performs better than other techniques in the literature, such as the ones summarized in \cite{Boyd_LMI_Book}, albeit with slightly higher computational cost.
Furthermore, the proposed method can yield a tighter polygon than the method considered in \cite{Siamak_Letter_BP} with similar computational cost, and can avoid a non-convex or unbounded polygon.

The organization of the paper is as follows: In Section \ref{Sec:Background}, background and definitions are given, the problem is stated, and a brief summary of existing techniques is provided.
The proposed method is described in more detail in Section \ref{Sec:Proposed}.
The performance analysis of each method in different scenarios is evaluated numerically in Section \ref{Sec:Simulations}.
Finally, Section \ref{Sec:Conclusion} concludes the paper.

\section{Background and Problem Statement}\label{Sec:Background}

\subsection{Notation}
Small and capital bold letters represent vectors and matrices, respectively.
The vector 2-norm operation is denoted by $\|\cdot\|$, and the matrix transpose and inverse operations are denoted by $(\cdot)^T$ and $(\cdot)^{-1}$, respectively.
The determinant of a matrix $\A$ is denoted by $\textsf{det}(\A)$.
The symbol $\I$ denotes an identity matrix of appropriate dimension.
The notation $\A \succ 0$ ($\A \prec 0 $) means that $\A$ is a positive definite (negative definite) matrix and $\A \succeq 0 $ ($\A \preceq 0$)  means that the matrix is symmetric positive semi-definite (negative semi-definite)
\cite{Horn}.
By $\x \in \mathbb{R}^M$ and $\X \in \mathbb{R}^{M \times N}$ we mean that the vector $\x$ and matrix $\X$ are of size $M$ and $M \times N$, respectively.

\subsection{Various Forms of Representing an Ellipsoid}\label{Subsec:Definition_Ellipsoids}
An ellipsoid in $\mathbb{R}^{\nu}$ can be defined by different but equivalent forms:
\begin{itemize}

\item Image of the unit ball: An ellipsoid can be obtained by mapping a unit ball as
\be \label{Unit_Ball_Transform}
\xi = \Big\{\x \in \mathbb{R}^{\nu}: \x = \P \y + \x_c, \| \y \| \leq 1 , \y \in \mathbb{R}^{\nu} \Big\}
\ee
where without loss of generality we can assume $\P \in \mathbb{R}^{\nu \times \nu}$ is symmetric positive definite, see, e.g., \cite{Modern_Convex_Optimization}.
The volume of the ellipsoid is $ V_{\nu} \textsf{det}( \P )$,  where $V_{\nu}$ is the volume of the unit ball in $\mathbb{R}^{\nu}$.

\item Quadratic form I:
\be \label{Quadratic_form_I}
\xi = \Big\{ \x \in \mathbb{R}^{\nu} : ~ \| \B \x + \d \| \leq 1 \Big\}
\ee
where $\B \in \mathbb{R}^{\nu \times \nu}$ and $\d \in \mathbb{R}^{\nu}$.
The two sets in \eqref{Unit_Ball_Transform} and \eqref{Quadratic_form_I} are identical when $\B = \P^{-1}$ and $\d = -\P^{-1} \x_c$.
Herein we also assume that $\B$ is symmetric positive definite.
The volume of the ellipsoid is then $ V_{\nu} \textsf{det}( \B^{-1} ) $.

\item Quadratic form II: 
\be \label{Quadratic_form_II}
\xi = \Big\{ \x \in \mathbb{R}^{\nu} : \x^T \A \x + 2 \x^T \b + c \leq 0 \Big\}
\ee
where $\A \in \mathbb{R}^{\nu \times \nu}$ is symmetric positive definite, $\b \in \mathbb{R}^{\nu}$, and $c \in \mathbb{R}$.
The sets in \eqref{Quadratic_form_I} and \eqref{Quadratic_form_II} are identical when $\A= \B^T\B$, $\b = \B^T\d$, $c = \d^T \d - 1$ (so $c=\b^T \A^{-1}\b-1$).
When $\A$, $\b$ and $c$ satisfy $c=\b^T\A^{-1}\b-1$, the volume of the ellipsoid is equal to $V_{\nu} \textsf{det}( \A^{-1/2} )$.

\end{itemize}

In this paper, by referring to an ellipsoid we mean the closed convex body in $\nu$-dimensions rather than just its boundary.
In 2-D, i.e, when $\nu=2$, the ellipsoid is referred to as an ellipse.
In this work, we will be using the above different forms of description for the same ellipse.

\subsection{Problem Statement and Existing Techniques} \label{Sec:Problem_Statement}

\subsubsection{Problem Statement}
We denote the intersection region of the ellipses
$\xi_i = \{ \x \in \mathbb{R}^2 :  \| \B_i \x + \d_i  \| \leq 1 \} $,
for $i=\{1,\ldots,M\}$, by 
\be \label{Ellipse_Basic}
{\cal E} = \bigcap_{i=1}^{M} \xi_i,
\ee
where without loss of generality we assume these ellipses are distinct.
Throughout this work, we refer to $\cal E$ as the \textit{feasible region}, and assume that it is a non-empty region.
The problem is to find the smallest area ellipse
$ \xi_0 = \{ \x \in \mathbb{R}^2: \| \B_0 \x + \d_0 \| \leq 1 \} $
such that it contains the feasible region, i.e., 

\be \label{Ellipsoid_Intersection_Region}
\xi_0 \supseteq {\cal E}.
\ee
Given $\xi_0$, verifying that \eqref{Ellipsoid_Intersection_Region} holds is NP-complete, and thus finding the smallest ellipse $\xi_0$ such that \eqref{Ellipsoid_Intersection_Region} holds is not tractable \cite[p.44]{Boyd_LMI_Book}.
However, there are some sub-optimal solutions to find less tight ellipsoids in arbitrary dimensions, which will be described below.

\subsubsection{Popular Existing Techniques} \label{Subsec:Existing_Techniques}
Different techniques have been proposed for finding a sub-optimal solution to the aforementioned problem.
Below, we describe two of the most popular techniques, each of which formulates the problem as a standard convex optimization one, which is solvable in polynomial time.

i) \textit{Approximation Using the Sufficient Condition}:
It is shown in \cite[p.44]{Boyd_LMI_Book} that by using the so called ${\cal S}$-procedure, the sufficient condition for \eqref{Ellipsoid_Intersection_Region} to hold can be expressed as a linear matrix inequality (LMI):
\be \label{LMI_Sufficient_Ellipsoid}
\bmx
\A_0 & \b_0 & 0 \\
\b_0^T & -1 & \b_0^T\\
0 & \b_0 & -\A_0 
\emx
-\sum_{i=1}^M
\tau_i
\bmx
\A_i & \b_i & 0\\
\b_i^T & c_i & 0 \\
0 & 0 & 0 
\emx
\preceq 0 , 
\ee
where $\tau_i$ for $i=\{1,\ldots,M \}$ are positive unknowns to be estimated, and $\A_i$, $\b_i$, and $c_i$  are related to $\B_i$ and $\d_i$ based on the definitions given earlier in Section \ref{Subsec:Definition_Ellipsoids}.
Note that $\A_0$, $\b_0$, and $c_0$ are normalized such that $c_0=\b_0^T\A_0^{-1}\b_0-1$.
This LMI is not a necessary condition for \eqref{Ellipsoid_Intersection_Region} to hold,
and thus with \eqref{LMI_Sufficient_Ellipsoid} we cannot characterize all the ellipses that cover the intersection of multiple ellipses.
However, among all the ellipses $\xi_0$ with variables $\A_0,\b_0$ satisfying \eqref{LMI_Sufficient_Ellipsoid}, one can find the best outer approximation of the intersection region of $\xi_1,\ldots,\xi_M$ by solving the following semi-definite programming (SDP) problem:
\be
\begin{split} \label{Ellipse_Outer_Technique_1}
& \min_{\A_0,\b_0,\tau_1,\ldots,\tau_M} ~ \log \det \A_0^{-1}  \\
& \textrm{subject to} \quad \A_0 \succ  0, ~ \tau_1,\ldots,\tau_M \geq 0, 
~ \eqref{LMI_Sufficient_Ellipsoid} \\
\end{split}
\ee

ii) \textit{Ellipse Obtained by Expansion of the Largest Inscribed Ellipse}:

Another approach \cite[p.414]{Boyd} is to first find the maximum area ellipse $\xi_{max} = \{ \x: \x= \P_0 \y + \x_{c_0} , \|\y \| \leq 1 \}$ inscribed by the intersection of several ellipses by solving
\begin{align} \label{Ellipse_John_Lowner}
&\max_{\P_0,\x_{c_0},\tau_1,\ldots,\tau_M} ~ \log \det \P_0 \nonumber \\
& \textrm{subject to} ~ 
\bmx
-\tau_i - c_i + \b_i^T\A_i^{-1}\b_i & \0 & (\x_{c_0}+ \A_i^{-1} \b_i )^T \\
\0 & \tau_i \I_{\nu} & \P_0 \\
\x_{c_0} + \A_i^{-1}\b_i & \P_0 & \A_i^{-1}
\emx \succeq 0,  \nonumber \\
&~ \quad \quad \quad \quad \quad \P_0 \succ0 ,~  \tau_i \geq 0, ~  \quad \quad  \quad \quad \quad \quad \quad \quad \quad i=1,\ldots,M
\end{align}
which is a convex SDP optimization problem whose solution can be obtained efficiently. 
Then as shown by John and Lowner, e.g., see \cite{John_Lowner} and the references therein, by scaling this ellipse by a factor of $\nu=2$ (since we are currently considering a 2-D space), an ellipse covering the intersection of multiple ellipses can be obtained, see \cite[p.414]{Boyd}.

Either of the above methods can be applied to find the ellipsoidal outer approximation of the intersection of ellipsoids with arbitrary dimensions.
However, due to the approximations made, the obtained ellipses might not always be tight (which will be shown through simulations), limiting their applicability.
In the following section, we develop a geometrical bounding method in 2-D space, which through numerical evaluation, is shown to offer a tighter outer approximation of the intersection of multiple ellipses.

\section{Proposed Bounding Ellipse Method}\label{Sec:Proposed}

Our solution to the problem consist of two stages: (i) finding a tight polygon, which contains $\cal E$, and (ii) finding the tightest ellipse, which contains the vertices of the obtained polygon. 
The second stage involves a well-known optimization problem, which can be solved by iterative optimization techniques, as done in \cite{Enclosing_6, Ellipse_Bounding_Points_2,Ellipse_Bounding_Points_3, Min_Vol_Ellipse_Points}.
It can also be formulated as a standard convex optimization problem and solved by standard optimization packages on a computer \cite[p. 411]{Boyd}.
The main aim of this paper is to find a tight polygon, with a small number of vertices, to cover $\cal E$ such that the smallest area ellipse which contains the vertices of the polygon is a tighter outer-approximation of $\cal E$ compared to the ellipses obtained by other approximation techniques mentioned in Section \ref{Subsec:Existing_Techniques}.
In the following, we describe the two main stages of our proposed technique in more detail.

\subsection{Finding a Tight Polygon Containing the Intersection of Ellipses}
We divide the task of finding a polygon containing $\cal E$ into several steps as follows:

\emph{Step 1 (Finding the intersection points of ellipses on the boundary of $\cal E$)}:
First, the intersection points of the boundaries of every pair of ellipses are found and the ones not lying in $\cal E$ are rejected.
Finding the intersection points of two ellipses can be done by computing the roots of a polynomial of degree 4, as discussed in \cite{Ellipse_Intersection_Points}.
The possible number of intersection points can vary from 0 to 4, and in the special case that the two ellipses are disks, this number could be 0, 1, or 2.
If the number of intersection points is 0 or 1, then either the two ellipses have no intersection region, or one of them is contained in the other one. 
Since we assume that $\cal E$ is a non-empty region, one of the ellipses has to contain the other. 
In this case we should remove the larger ellipse in the process of finding a tight polygon.
To this end, we generate one point on the boundary of each ellipse randomly, and if one of these points does not satisfy the defining inequality of the other ellipse, then the former ellipse is the larger one.
In the worst case there are 4 intersection points for every pair of ellipses and since there are $M(M-1)/2$ different pairs of ellipses, there will be at most $2M(M-1)$ intersection points to be verified. 
Since checking if an intersection point satisfies the inequalities of the remaining $M-2$ ellipses takes ${\cal O} (M )$ operations, ${\cal O}(M^3)$ operations are sufficient to find the intersection points on the boundary of $\cal E$. 
The ${\cal O} (M^3 )$ algorithm implemented herein is straightforward and can easily be implemented. It is also very efficient when the value of $M$ is small. Theoretically faster algorithms exist that run closer to ${\cal O} (M^2)$ if the arrangement of the ellipses is computed [31]. For very large values of $M$ such algorithms may turn out to be useful. However, they are also computationally involved, and it is not clear if they are useful in practice.

There is a possibility that some of the intersection points are non-distinct, e.g., more than two ellipses intersect at exactly the same point.
Note that due to rounding errors, two nearby points might also be regarded as the same points. This does not result in error, however, the obtained polygon might be slightly less tight if one of these points are used in the algorithm.
%
In these cases, we only use one of these points in our algorithm but keep the indices of the ellipses corresponding to these intersection points. 
For later use, the total number of intersection points remaining on the boundary of $\cal E$ is denoted by $m_{c}$.

\begin{figure}[tbp]
\centering
\includegraphics[width=95mm,height=75mm]{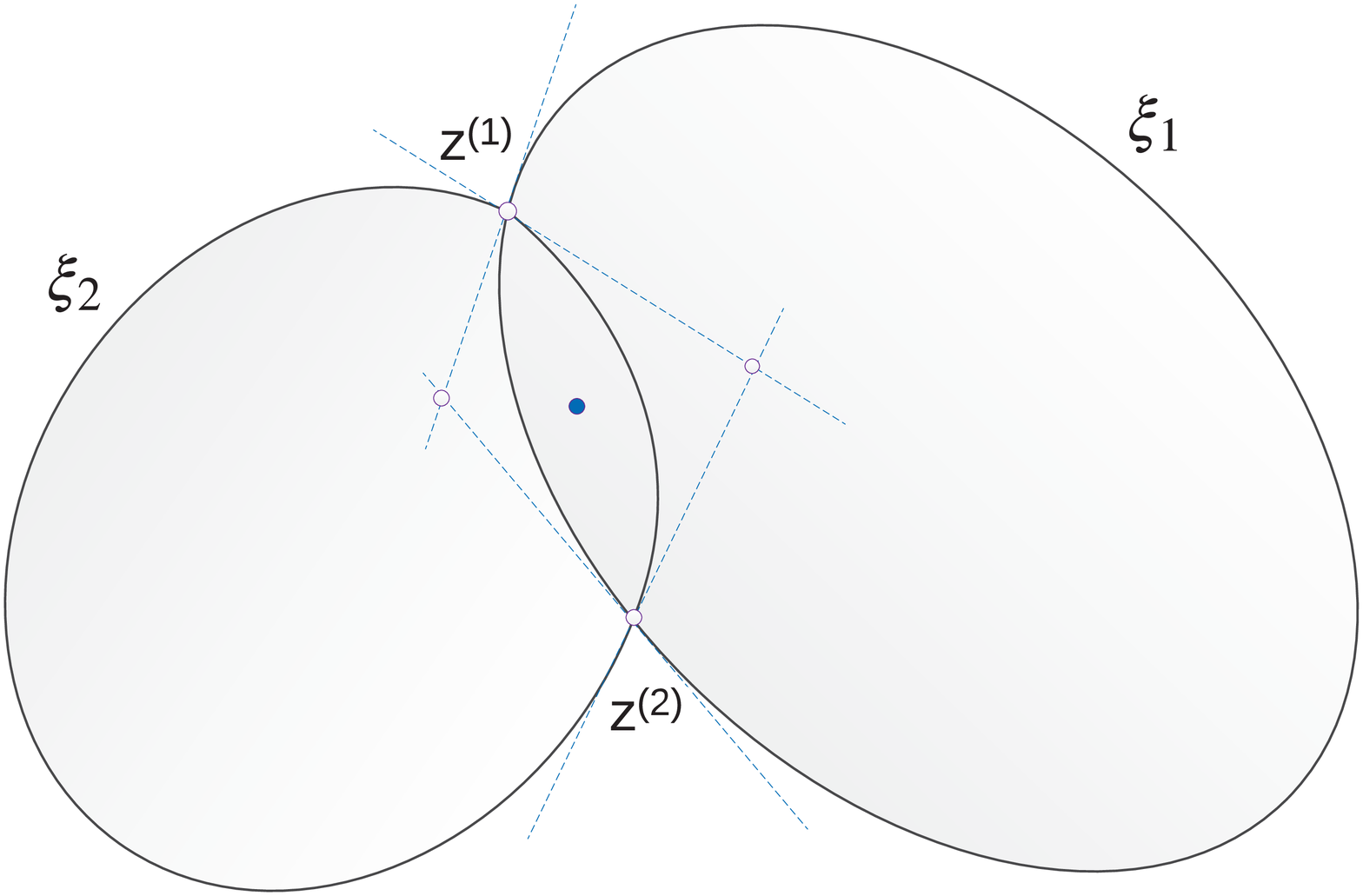}
\caption{The intersection of the half-planes tangent at the intersection points of the boundary of the ellipses forms a closed polygon.}
\label{fig:Intersect_Ellipse_Only}
\end{figure}

\emph{Step 2 (Generating extra points on $\cal E$)}:
After rejecting the intersection points not on the boundary of $\cal E$, we let $z^{(l)}$ for $l=1,\ldots, m_{c}$ be the remaining intersection points.
We find the mean of the these intersection points as
\be
\z_{\textrm{mean}} = \frac{1}{ m_c} \sum_{ l \in m_c} \z^{(l)}
\ee
The vectors connecting $\z_{\textrm{mean}}$ to the intersection points are 
\be 
\label{Vector_Average_Point}
\v^{ (l) } = \z^{(l)} - \z_{ \textrm{mean} } , \quad l = 1,\ldots, m_c  
\ee
and the angles they make with the x-axis in the Cartesian coordinates are denoted by $\alpha^{(l)} \in [0,2\pi)$.
The vectors $\v^{(l)}$ are then sorted according to increasing angles.
The intersection points may be used to generate a polygon covering $\cal E$ as shown in Fig. \ref{fig:Intersect_Ellipse_Only}.
However, as observed, the generated polygon may not be tight enough if only the intersection points are used.
Furthermore, we might face degeneracy problems (to be illustrated with examples) where the intersection of half-planes forms an unbounded polygon which can not be used as a finite outer-approximation of the feasible region. 
To overcome these problems, we generate a number of additional points on the elliptic arc segments of $\cal E$ between  two neighbouring intersection points.

\begin{figure}[tp]
\centering
\includegraphics[width=130mm,height=60mm]{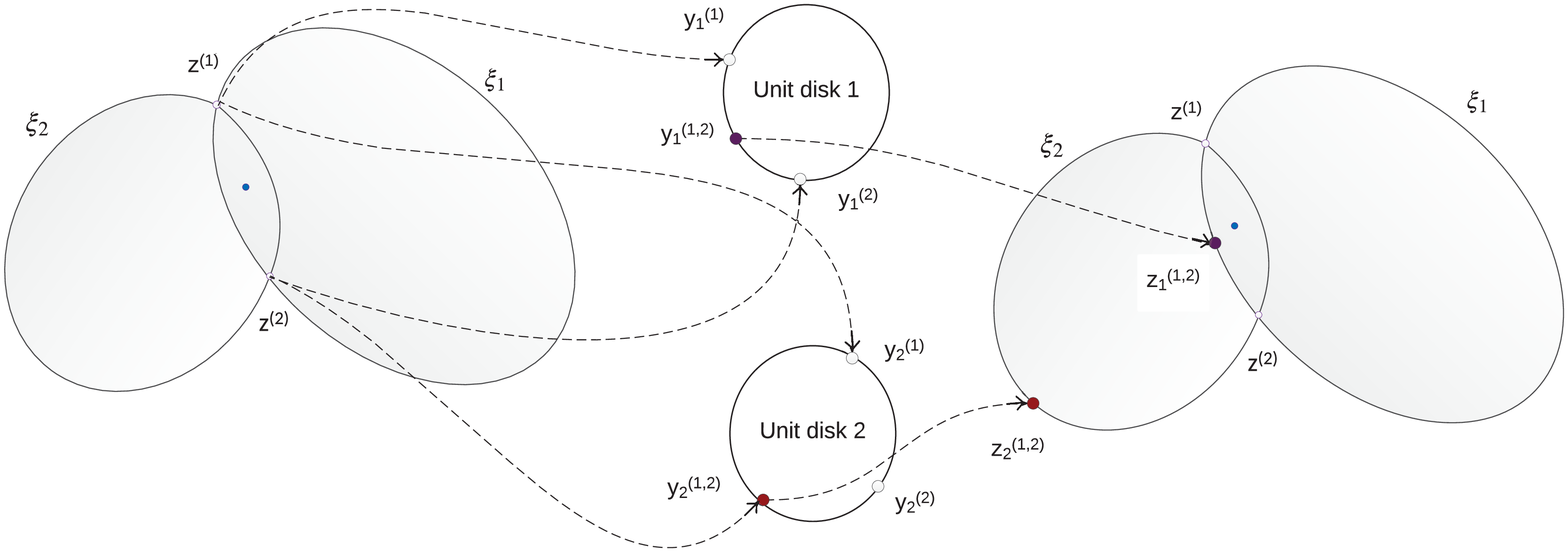}
\caption{Detecting the curve connecting $\z^{(1)}$ to $\z^{(2)}$ in a counter-clockwise manner.}
\label{fig:Ellipse_Disk_Mapping}
\end{figure}

\begin{figure*}[htbp]
\centering
\subfloat[]{
\includegraphics[width=60mm,height=55mm]{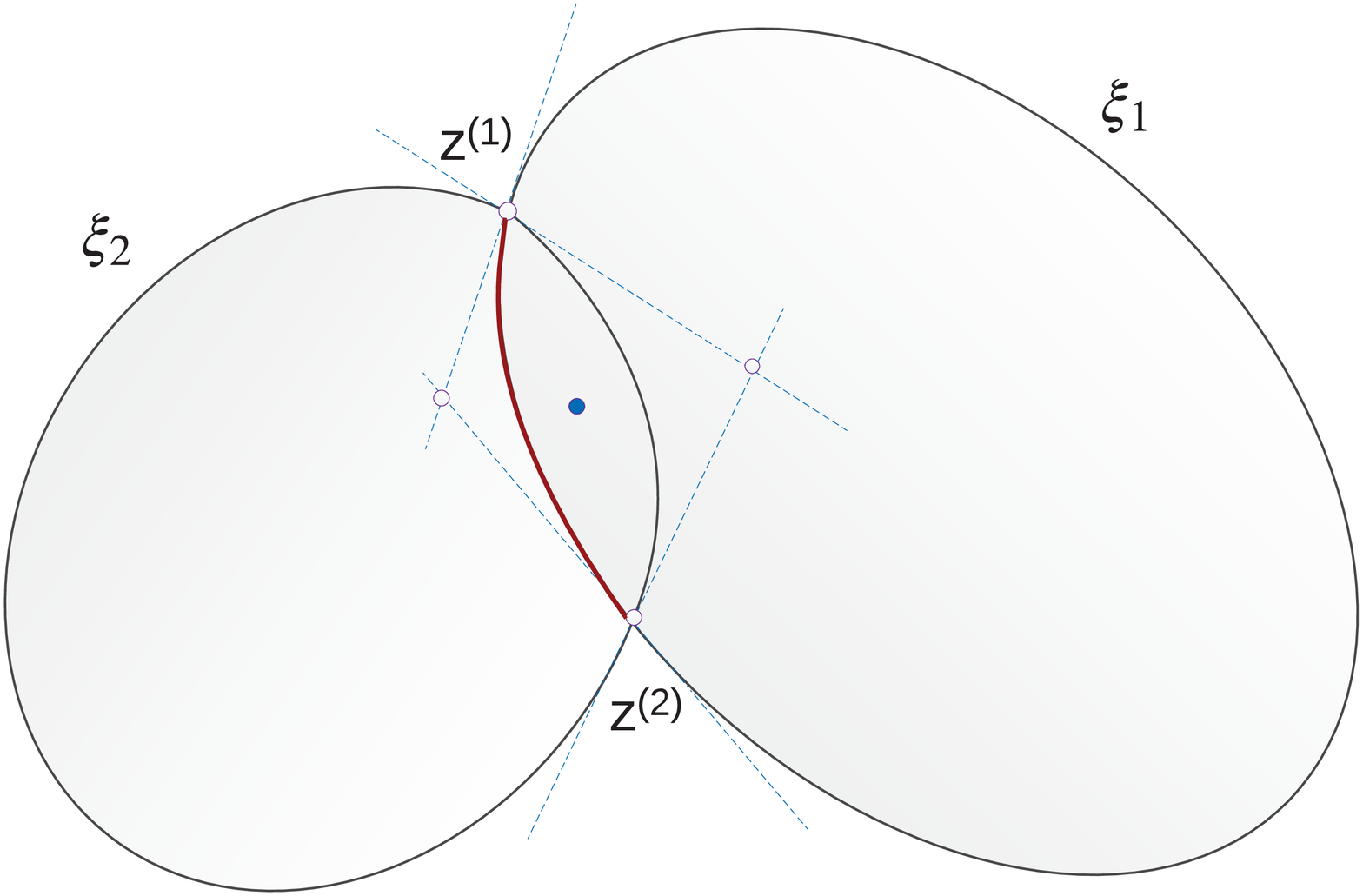}
}
\subfloat[]{
\includegraphics[width=60mm,height=55mm]{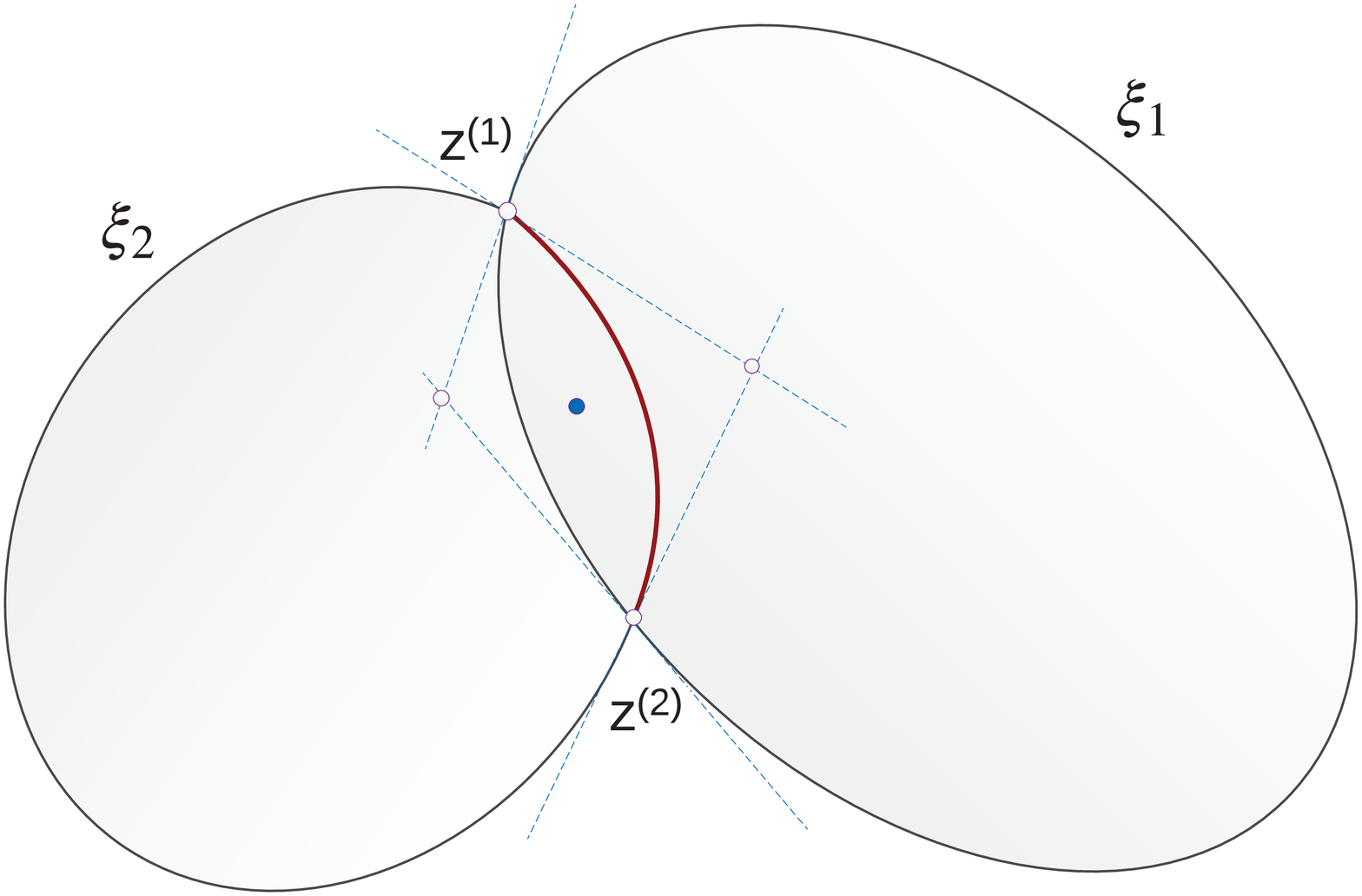}
}
\caption{The detected arc segments and the corresponding ellipses forming $\cal E$. (a) The segment (shown with red) is detected to be from $\xi_1$, connecting $\z^{(1)}$ and $\z^{(2)}$. (b) The segment (shown with red) is from $\xi_2$, connecting $\z^{(2)}$ and $\z^{(1)}$.}
\label{fig:Intersect_Ellipse_Segment}
\end{figure*}

To do this, we need to know the ellipse corresponding to each elliptic arc segment forming the boundary of $\cal E$.
Note however that as mentioned earlier, an intersection point $\z^{(l)}$ could be related to more than two ellipses.
Thus, in general, the two neighbouring points correspond to several different ellipses (minimum of two different ones), among which some are common.
Obviously, the elliptic arc segment on the boundary of $\cal E$ that connects the two intersection points corresponds to an ellipse, the boundary of which passes through both neighbouring points. 
Thus the index of this ellipse is common to both intersection points.

Hence we remove the indices of the ellipses that are not common to both neighbouring intersection points, and assume that the remaining indices form a set, temporarily denoted as ${\cal P}$.
With two or more indices left, there are multiple elliptic arc segments that connect the two intersection points, among which only one of them is part of the boundary of ${\cal E}$; thus there is an ambiguity in knowing the ellipse corresponding to that arc segment. 
To resolve this ambiguity and detect the correct indices, we randomly generate one point on the boundary of each ellipse, with index $k \in {\cal P}$, between the two intersection points (in a counter-clockwise manner) and see which point satisfies all the inequalities of the remaining ellipses, i.e., it falls inside or on the boundary of all the ellipses, whose indices are in ${\cal P}$.
Specifically, to generate a point randomly on each ellipse, the two neighbouring intersection points, $\z^{(l_c)}$ and $\z^{(l_c+1)}$ where $l_c \in \{1,\ldots, m_c-1 \}$, are mapped onto the boundary of unit disks based on the equations of the corresponding ellipses, i.e., the inverse mapping defined in \eqref{Unit_Ball_Transform}, as
\begin{align} \label{Transform_Unit_CaseI_1}
\y_k^{(1)} &= \P_k^{-1}(\z^{(l_c)} - \x_{c,k}) \\
\label{Transform_Unit_CaseI_2}
\y_k^{(2)} &= \P_k^{-1}(\z^{(l_c+1)} - \x_{c,k})  
\end{align}
Assume that the two generated points on the $k$-th disk are denoted by $\y_k^{(1)}$ and $\y_k^{(2)}$ and the angles corresponding to these points are denoted by $\theta_k^{(1)}$ and $\theta_k^{(2)}$, respectively.
Now we generate one point on the arc corresponding to the $k$-th disk connecting $\y_k^{(1)}$ and $\y_k^{(2)}$ and then transform this point $\y_k^{(1,2)}$ back onto the $k$-th ellipse by means of:
\be \label{Circle_to_Ellipse_2}
\z_k^{(1,2)} = \P_k \y_k^{(1,2)} + \x_{c,k},
\ee
Then by verifying if $\z_k^{(1,2)}$ falls inside or on the boundary of all the remaining ellipses with indices in $ {\cal P}$, we can determine the ellipse which is forming the boundary of ${\cal E}$.
%
The actions done so far in \emph{Step 2} are summarized in Algorithm I, lines 14-23.

The process of detecting the ellipses which form ${\cal E}$ is shown with a simple example in Fig. \ref{fig:Ellipse_Disk_Mapping}.
The two ellipses, labeled as $\xi_1$ and $\xi_2$, have two intersection points $\z^{(1)}$ and $\z^{(2)}$, which are first ordered based on the angles the corresponding vectors $\nu^{(l)}$ \eqref{Vector_Average_Point} make with the \textit{x}-axis. 
We start from the point with the smallest angle, i.e., $\z^{(1)}$ and try to find the curve connecting it to the neighbouring point with the next smallest angle, which is $\z^{(2)}$, as shown in Fig. \ref{fig:Ellipse_Disk_Mapping}. 
The two points are mapped onto the boundary of unit disks based on the equations of each ellipse, yielding, $\y_1^{(1)}$ and $\y_1^{(2)}$ for the disk corresponding to $\xi_1$ and $\y_2^{(1)}$ and $\y_2^{(2)}$ for the disk corresponding to $\xi_2$. 
Then a point is generated on the boundary of each disk on the arc connecting the two mapped points in a counterclockwise manner, yielding $\y_1^{(1,2)}$ and $\y_2^{(1,2)}$.
The obtained points, $\y_1^{(1,2)}$ and $\y_2^{(1,2)}$ are then mapped back onto the ellipse corresponding to each disk, yielding $\z_1^{(1,2)}$ and $\z_2^{(1,2)}$, respectively.
Since $\z_1^{(1,2)}$ on $\xi_1$ is the point which falls on $\cal E$, the corresponding elliptic arc segment, shown in red in Fig. \ref{fig:Intersect_Ellipse_Segment}-(a), is detected to be the one sought after, and $\xi_1$ is the corresponding ellipse. The same process is repeated for detecting the elliptic arc connecting $\z^{(2)}$ to $\z^{(1)}$ in a counter clockwise manner, which becomes the one corresponding to $\xi_2$, as shown in red in Fig. \ref{fig:Intersect_Ellipse_Segment}-(b).

Suppose that the $j$-th ellipse remains after removing all the indices of the irrelevant ellipses.
Then the next task is to generate a number of points on the elliptic arc segment of the $j$-th ellipse, between the two neighbouring intersection points, e.g., $\z^{(l_c)}$ and $\z^{(l_c+1)}$, where $l_c \in \{1,\ldots, m_c-1 \}$.
To do so, we first generate a certain number of points on the boundary of the unit disk obtained by inverse mapping of the ellipse $\xi_j$, between the points $\y_j^{(1)}$ and $\y_j^{(2)}$ in a counter-clockwise manner.
The angles between a reference axis and the vectors connecting these points $\y_j^{(1)}$ and $\y_j^{(2)}$ to the origin, are computed and denoted as $\theta_j^{(1)}$ and $\theta_j^{(2)}$, respectively.
Depending on the difference between $\theta_j^{(1)}$ and $\theta_j^{(2)}$ we can generate a number of points on the unit circle.
For instance, if in total it is desired to generate $m$ points on each circle, the arc length between every two points is $2\pi/m$.
Thus we may want to generate the points $\y_{1,2}^{(l_j)}$ between $\y_j^{(1)}$ and $\y_j^{(2)}$ on the unit circle, for $l_j=1,\ldots,\texttt{floor}[(\theta_j^{(1)} - \theta_j^{(2)} )m/(2\pi)]$, where $\texttt{floor[.]}$ returns the largest integer not greater than its argument.
After generating the points $\y_{1,2}^{(l_j)}$ on the unit circle, they are transformed back onto the $j$-th ellipse by means of 
\be \label{Circle_to_Ellipse}
\z_j^{(l_j)} = \P_j \y_{1,2}^{(l_j)} + \x_{c,j}, \quad l_j = 1, \ldots, \texttt{floor}[(\theta^{(1)} - \theta^{(2)} )m/(2\pi )]
\ee
The half planes, tangent to the $j$-th ellipse at the corresponding generated points are computed as follows
\be
(\B_j \z_{j}^{(l_j)} + \d_j)^T ( \B_j \x + \d_j)  \leq 1, \quad l_j = 1, \ldots, \texttt{floor}[(\theta^{(1)} - \theta^{(2)} )m/(2\pi )]
\ee
and the half planes, tangent to the $j$-th ellipse at the two ends of the arc segment under consideration, i.e., $\z^{(l_c)}$ and $\z^{(l_c+1)}$ are calculated as
\be
(\B_j \z^{(l)} + \d_j)^T ( \B_j \x + \d_j)  \leq 1, \quad l=l_c, l_c +1
\ee
This process of generating point using mapping on to unit disk is summarized in Algorithm I, lines 24-28.

In this way, for each arc of the feasible region connecting the two intersection points, a number of discrete points, denoted by $m_d$, are generated and the half-planes, tangent to the ellipse corresponding to that segment (e.g., $j$-th ellipse) are computed.
For the $m_c$ intersection points also $2m_c$ half planes are obtained.
Therefore, in total there will be $m_d+m_c$ discrete points with $m_d + 2m_c$ half planes corresponding to these points.

\emph{Step 3 (Intersecting the tangent lines to find the vertices of the polygon)}:
The intersection of the obtained $m_d+2m_c$ half planes usually forms a bounded and convex polygon.
One way to compute the vertices of the polygon is to use the divide-and-conquer algorithm \cite{Half_Plane_Intersection_1}, or simplified versions thereof using the techniques proposed in \cite{Polygon_Intersection_1, Polygon_Intersection_2}; in which the cost is
${\cal O} \Big( (m_d+m_c) \log(m_d+m_c) \Big)$.
However, we will find the polygon more efficiently for the 2-D case as explained below. Since $\cal E$ is convex and its boundary is piecewise smooth, it can be observed that each vertex of the bounding polygon is the intersection of two tangent lines corresponding to two neighbouring points.
Thus to obtain these vertices, we solve the linear system of equations corresponding to the two tangent lines at two neighbouring points on each segment.
Since we only need to solve $m_d+m_c$ systems of linear equations (the intersection of tangent lines at the intersection points of ellipses are already obtained), only ${\cal O }(m_d+m_c)$ flops are sufficient.
This step is summarized in lines 24-29 of Algorithm I.

\begin{figure*}[tbp]
\centering
\subfloat[]{
\includegraphics[width=55mm,height=55mm]{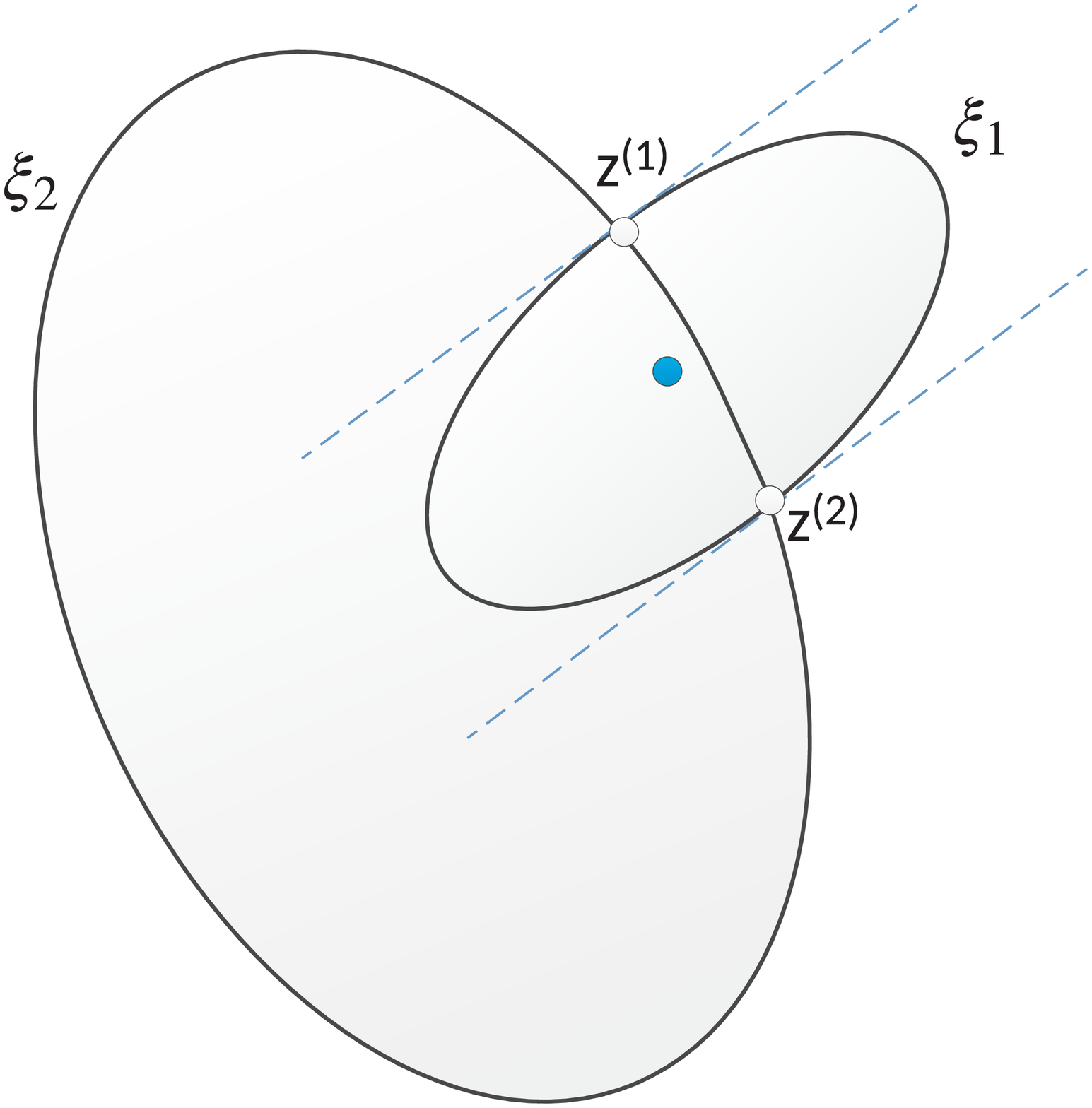}
}
\subfloat[]{
\includegraphics[width=55mm,height=55mm]{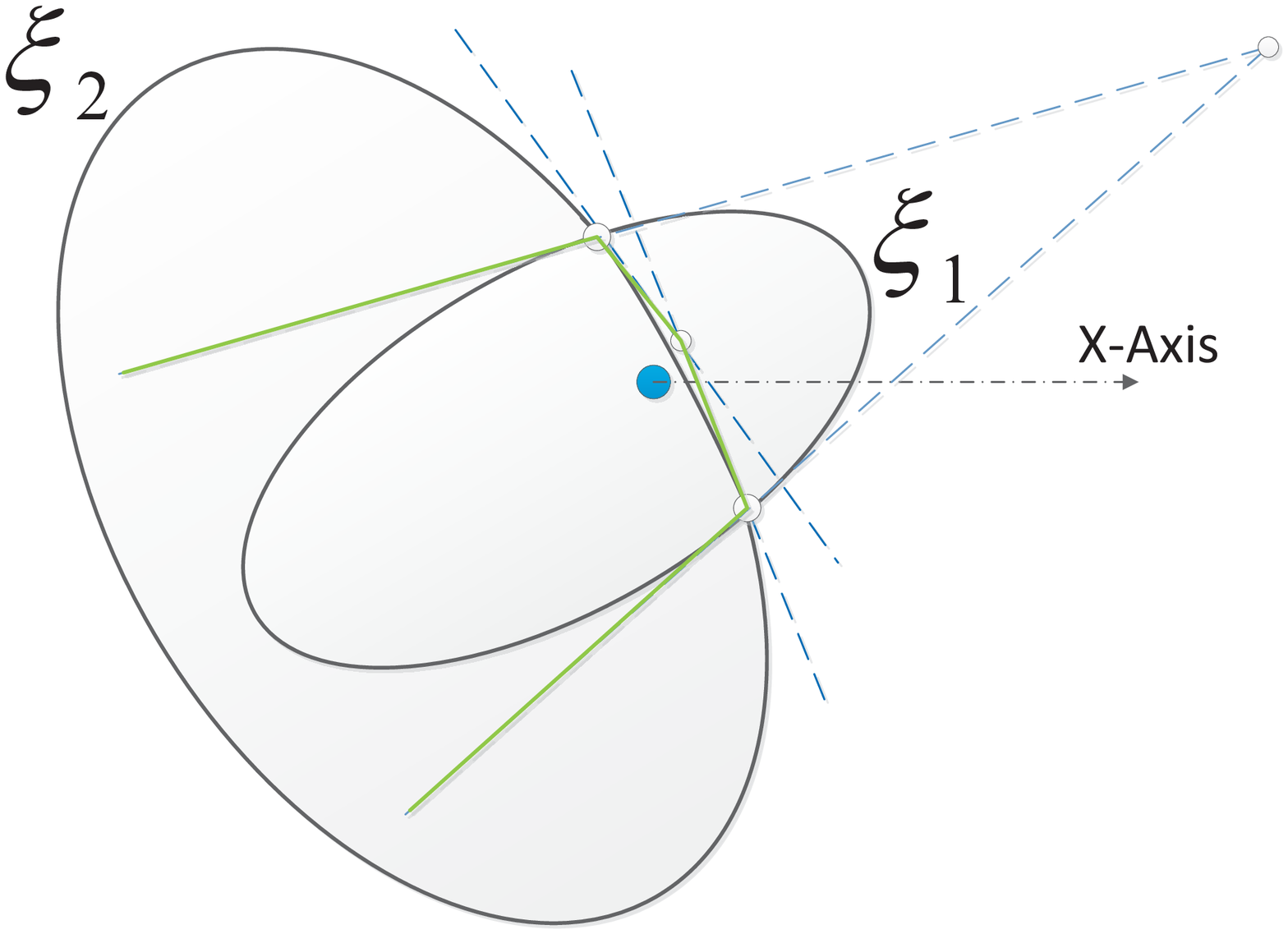}
}
\caption{The degenerate cases where a closed polygon can not be obtained: (a) parallel tangent lines, (b) the intersection point does not satisfy the affine inequalities corresponding to the other half-planes.}
\label{fig:Degenracy_1}
\end{figure*}

\emph{Step 4 (Detecting the degeneracy problem)}:
Even after the above steps, there is a possibility of degeneracy problems, i.e., the polygon formed is not bounded or convex and does not cover $\cal E$.
These two situations, illustrated in Fig. \ref{fig:Degenracy_1}, are as follows:

\begin{itemize}
\item If the tangent lines corresponding to two neighbouring points do not intersect, i.e., the two lines are parallel, the polygon will be unbounded.
Therefore, the number of points on the elliptic arc connecting two neighbouring points needs to be increased.
\item If the intersection exists but does not satisfy the remaining of the $m_d+2m_c$ affine inequalities, this point can not be a vertex of the desired polygon.
Note that the desired polygon should be formed as a result of intersecting half-planes, and if the intersection region is unbounded, a bounded convex polygon which contains $\cal E$ can not be found by relying only on these half-planes.
Consequently, the tangent lines corresponding to two neighbouring points cannot be the sides of the polygon, and thus there are no support lines to $\cal E$ at these points. 
\end{itemize}

Detecting the first case is simple as the intersection of two parallel lines has no solution.
To detect the second case, we find the intersection of the tangent lines corresponding to every two neighbouring points and verify if it satisfies all the affine inequalities corresponding to the half planes of the other discrete points.
This can be done with ${\cal O}(m_d+m_c)$ flops as there are $m_d+2m_c-2$ remaining affine inequalities to be verified for every obtained point. 
Detecting the degeneracy can be done for each arc segment just after \emph{Step 3} is implemented for that segment.
If degeneracy occurs, then we only increase the number of points on the specific arc segment by generating $\texttt{floor}[(\theta^{(1)} - \theta^{(2)} )(m+\Delta m)/(2\pi )]$ points on the unit circle and then mapping them onto the corresponding ellipse, where $\Delta m$ is a relatively small integer.
One strategy is to always double the value of $m$ so it can guarantee quickly finding a number that can avoid degeneracy.

At the end of these steps, the intersection of every pair of tangent lines corresponding to two neighbouring points, as well as the previously obtained intersection points of every pair of ellipses, represent the $m_p$ vertices of the desired polygon.\footnote{Note that the generated intersection points on the boundary of $\cal E$, excluding the intersection points of ellipses, are not required to represent the polygon because they lie on its sides.}

\subsection{Finding the Tightest Ellipse Containing the Polygon}
Assume that we have found a relatively tight polygon, represented by vertices $\tilde{\z}^{(l)}$ for $l = 1 , \ldots, m_p$, which covers $\cal E$.
Then one can find the smallest area ellipse (minimum spanning ellipse) which contains the vertices of this polygon (and hence contains $\cal E$) as done in \cite{Min_Vol_Ellipse_Points, Enclosing_6, Ellipse_Bounding_Points_2, Ellipse_Bounding_Points_3}.
This problem can also be formulated as a standard convex optimization problem:
\begin{align} \label{Min_Vol_Ellipse_Points}
& \min_{\B_0, \d_0} ~\log \det \B_0^{-1}  \nonumber \\
& \textrm{subject to}  ~\| \B_0 \tilde{\z}^{(l)} + \d_0 \| \leq 1 ,\quad l = 1,\ldots, m_p
\end{align}
where $\log \textrm{det}  \B_0^{-1}$ is proportional to the area of the ellipse \cite{Boyd}.
Since the inequalities can also be written as an LMI, this optimization problem can be formulated as a standard SDP. 
If the feasible region is tightly outer-bounded by the polygon, then it is very likely that it is tightly outer-approximated by the bounding ellipse.

\subsection{Algorithm Summary and Remarks}
The proposed method is summarized in Algorithm \ref{Algorithm_Method_II}.
Sometimes, one is interested in obtaining a desired bounding ellipse with a certain level of tightness, and may not know in advance how to chose an appropriate $m$.
To make the algorithm useful for such applications, after solving \eqref{Min_Vol_Ellipse_Points}, we can increase the number of points by generating one point between every two points that we have already generated.
Then we can compare the area of the updated bounding ellipse with the one corresponding to the previous value of $m$.
If their difference is more than a predefined threshold then we continue this process.
If for the initial $m$, a bounded polygon cannot be formed, we increase $m$ until a bounded polygon is formed, and then proceed as described above.
By iterating this process, we can determine if the area of the bounding ellipse obtained by the proposed technique has converged to some limit, and thus a possibly tight ellipse is obtained.

\begin{algorithm}[tbp]
\caption{Ellipse Outer-approximation}
\begin{algorithmic}[1]  \label{Algorithm_Method_II}
\STATE Set a predefined maximum iteration number $k_{max}$
\FOR { $k_{\textrm{Iter}} = 1$ to $k_{\textrm{max}}$ }
\FOR {every pair of ellipses}
\STATE Find the intersection points of the ellipses.
\IF {the number of intersection points is 0 or 1}
\STATE Find the ellipse containing the other one and remove it from the set of intersecting ellipses.
\ENDIF
\IF {the number of remaining ellipses is equal to 1}
\STATE Use that ellipse as the tightest outer approximation of the feasible region.
\STATE End the algorithm.
\ENDIF
\ENDFOR
\STATE Reject the intersection points not lying on the boundary of $\cal E$ and get $\z^{(l)}$ for $l=1,\ldots,m_c$.
\STATE Find the mean of the intersection points and find the vector connecting the former to the latter.
\STATE Find the angles between the x-axis and these vectors and sort them according to their angles.
\FOR {every two neighbour points}
\STATE Remove the indices of the ellipses not common between two neighbouring intersection points.
\IF { $ | {\cal P} | \geq 2 $}
\STATE For each ellipse with index $k \in {\cal P}$, map the points onto the corresponding unit circle.
\STATE For $\y_k^{(1)}$ and $\y_k^{(2)}$, find the angles between the reference axis and the vectors connecting them to the centre of the unit circle, i.e., $\theta_k^{(1)}$ and $\theta_k^{(2)}$, respectively.
\STATE For every pair of points, generate one point on the unit circle with angle between $\theta_k^{(1)}$ and $\theta_k^{(2)}$.
\STATE Map every point back onto the corresponding ellipse and verify if it lies on the feasible region.
\ENDIF
\STATE Denote the remaining ellipse with index $j$.
\STATE Map the two neighbouring points $\z^{(l_c)}$ and $\z^{(l_c+1)}$ onto a unit circle through \eqref{Transform_Unit_CaseI_1} and \eqref{Transform_Unit_CaseI_2}.
\STATE Find the angles between the Cartesian axis and the vector connecting $\y_j^{(1)}$ and $\y_j^{(2)}$ to the centre of the unit circle, i.e., $\theta_j^{(1)}$ and $\theta_j^{(2)}$, respectively.
\STATE Generate $\texttt{floor}[(\theta_j^{(1)}-\theta_j^{(2)} )m/(2\pi)] $ points on the curve between $\y_j^{(1)}$ and $\y_j^{(2)}$.
\STATE Map the points back onto the ellipse and find the tangent lines to the curve at those points.
\STATE Find the intersection of every two neighbouring points to obtain the vertices of the polygon $\tilde{z}^{(l)}$.
\STATE Do the test in \emph{Step 4} to check if a degenerate case has happened.
\IF  {degeneracy occurs}
\STATE  $m \leftarrow m+\Delta m$ and go to line (27).
\ENDIF
\ENDFOR
\ENDFOR
\end{algorithmic}
\end{algorithm}

\section{Numerical Results} \label{Sec:Simulations}
In this section, we compare the performance of different algorithms in obtaining an ellipse to cover the intersection of several ellipses.
To this end, we use Matlab 2010b on a 64-bit computer with Intel i7-2600 3.4GHz processor and 12GB of RAM.
The metric utilized for evaluation is the area of the bounding ellipse $a_E$, which based on the different formats given in Section \ref{Subsec:Definition_Ellipsoids}, is equal to $a_{\nu} \textsf{det}(\P)$, $a_{\nu} \textsf{det}(\B^{-1})$, or $a_{\nu} \textsf{det}(\A^{-1/2})$, where $a_{\nu}=\pi$ is the area of the unit disk in 2D.
We randomly generate $M=2$, $M=3$, $M=5$, $M=10$, and $M=20$ intersecting ellipses to see the performance of our technique in different situations.
%
For comparison, we solve the optimization problem in \eqref{Ellipse_Outer_Technique_1} and denote this method as \emph{S-procedure}.
We also consider finding the ellipse by solving \eqref{Ellipse_John_Lowner} and then expanding it by a factor of $\nu=2$ to find an ellipse covering the intersection region. This method is denoted as \emph{Expanded} in the rest of this paper.
We also consider the technique proposed in \cite{Siamak_Letter_BP} where a bounding polygon is obtained in a different way than the method proposed in this paper. 
However, the optimization problem in \eqref{Min_Vol_Ellipse_Points} is finally solved to find the smallest area ellipse covering the polygon. This method is denoted as \emph{YWCC16} throughout this section.
For solving the optimization problems we use Sedumi solver \cite{SeDuMi} and CVX optimization toolbox for Matlab \cite{CVX_Grant_Boyd}.
We compare the performance of our proposed method, denoted as \emph{Proposed}, with the aforementioned approaches.

The areas of the obtained bounding ellipses, as well as the computation times for each method are given below for different scenarios.
The \emph{Expanded} and \emph{S-Procedure} methods do not depend on $m$ (i.e., the number of discrete points generated on each ellipse in \emph{YWCC16} and \emph{Proposed}), thus their performances do not change.
However, \emph{YWCC16} and \emph{Proposed} depend on $m$ and in general the area of the obtained bounding ellipse decreases with $m$.
We also compare the computation time of each method for the corresponding values of $m$ and see which one generates a tighter ellipse with a lower computational cost.
Since evaluating the computational cost in terms of number of operations is exhaustive and difficult, we use CPU time as a metric for comparison of computational cost.

\begin{figure*}[h]
\centering
\subfloat[]{
\includegraphics[width=55mm,height=45mm]{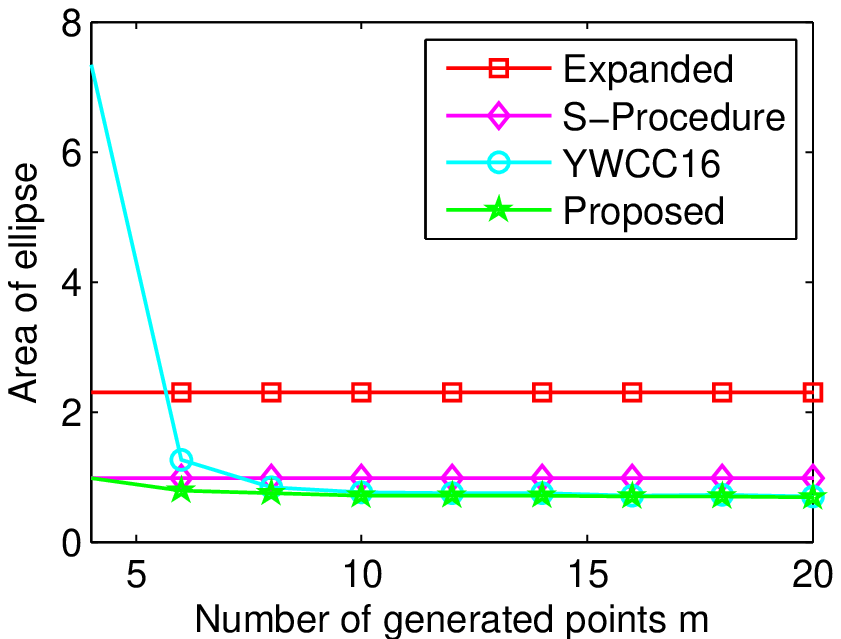}
}
\subfloat[]{
\includegraphics[width=55mm,height=45mm]{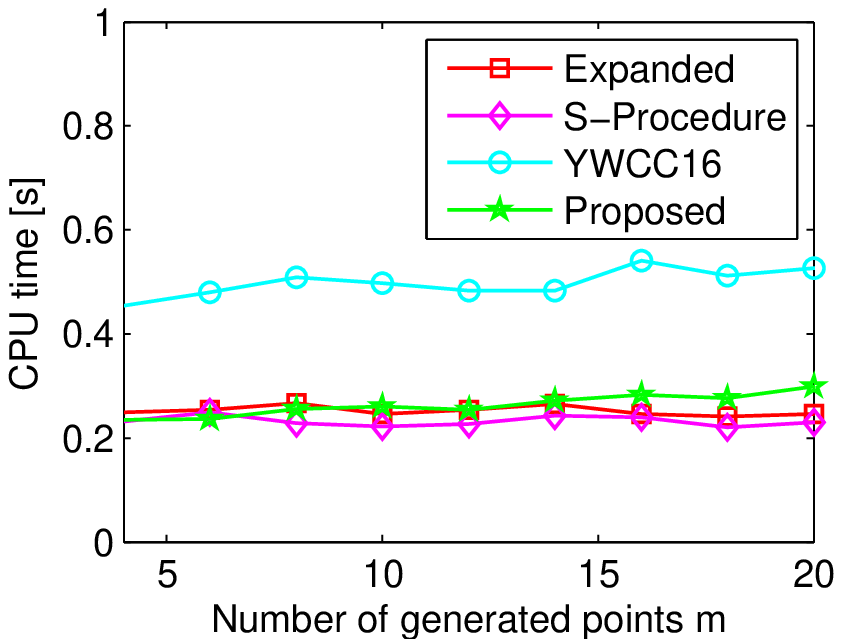}
}
\caption{Numerical test for M=2: (a) Areas of bounding ellipses, (b) CPU times.}
\label{fig:Ellipse_2}
\end{figure*}

Fig. \ref{fig:Ellipse_2}-(a) illustrates the area of the obtained ellipse for the scenario of $M=2$ ellipses.
As observed, the areas obtained by \emph{YWCC16} and \emph{Proposed} are slightly larger than the one obtained by \emph{S-Procedure}, however by increasing $m$, they can obtain a tighter ellipse compared to \emph{S-Procedure}.
The \emph{Expanded-inner} has the worst performance among all with area of 2.5 units.
The computation times of the different methods are given in Fig. \ref{fig:Ellipse_2}-(b).
As observed, \emph{Proposed} has similar computation time to \emph{S-Procedure} and \emph{Expanded-inner}, while \emph{YWCC16} has slightly higher computational cost. Increasing the number of initial discrete points $m$ does not change the computation times of \emph{YWCC16} and \emph{Proposed} noticeably.

\begin{figure*}[h]
\centering
\subfloat[]{
\includegraphics[width=55mm,height=45mm]{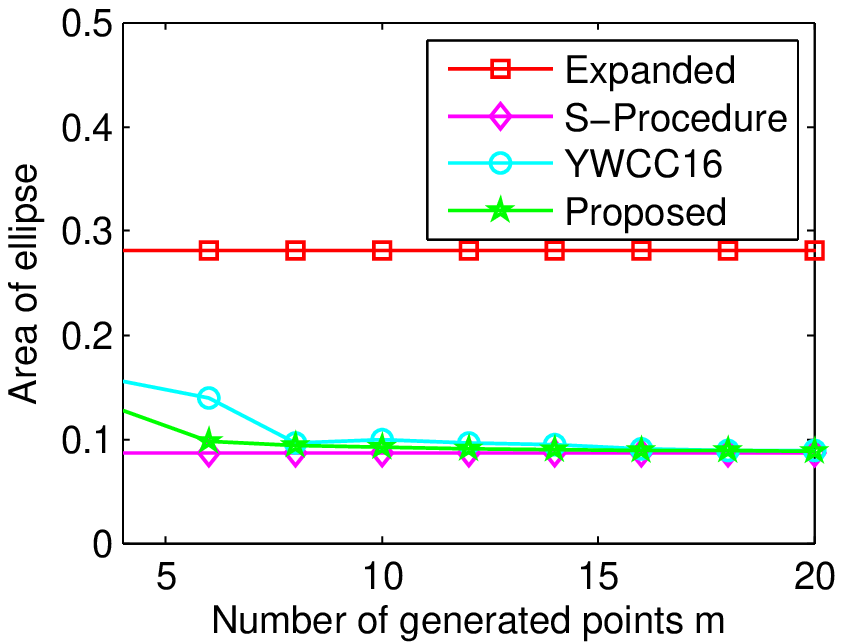}
}
\subfloat[]{
\includegraphics[width=55mm,height=45mm]{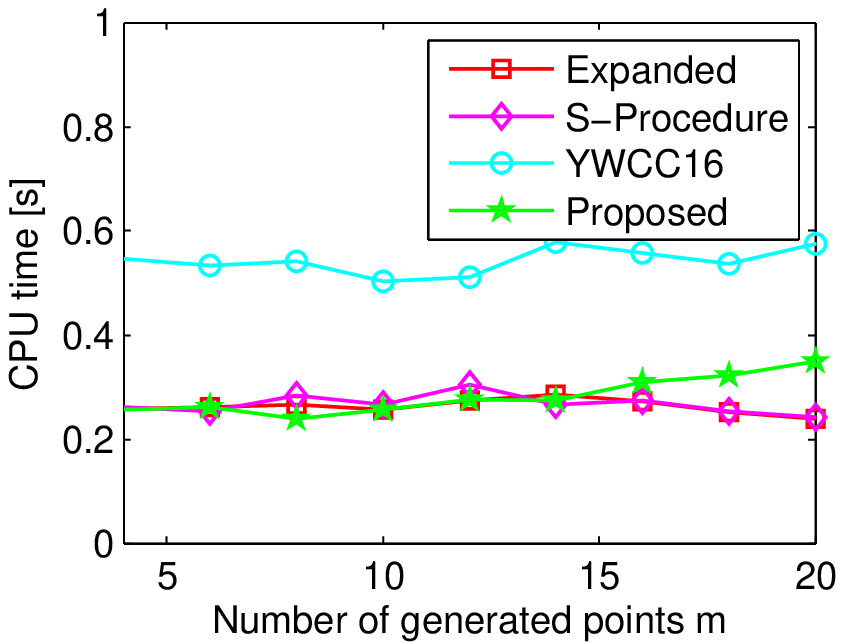}
}
\caption{Numerical test for M=3: (a) Areas of bounding ellipses, (b) CPU times.}
\label{fig:Ellipse_3}
\end{figure*}

In Fig. \ref{fig:Ellipse_3}-(a), we compare the area obtained by different methods for the scenario in which three ellipses intersect, i.e., $M=3$.
As observed, areas of the ellipses obtained by \emph{Proposed} and \emph{YWCC16} converge to the one obtained by \emph{S-Procedure} by increasing $m$, while \emph{Expanded} has the worst performance.
The computation times of different techniques are also given in Fig. \ref{fig:Ellipse_3}-(b). 
The CPU time for \emph{Proposed}, \emph{Expanded} and \emph{S-Procedure} are almost the same while \emph{YWCC16} has higher run-time.
%

\begin{figure*}[h]
\centering
\subfloat[]{
\includegraphics[width=55mm,height=45mm]{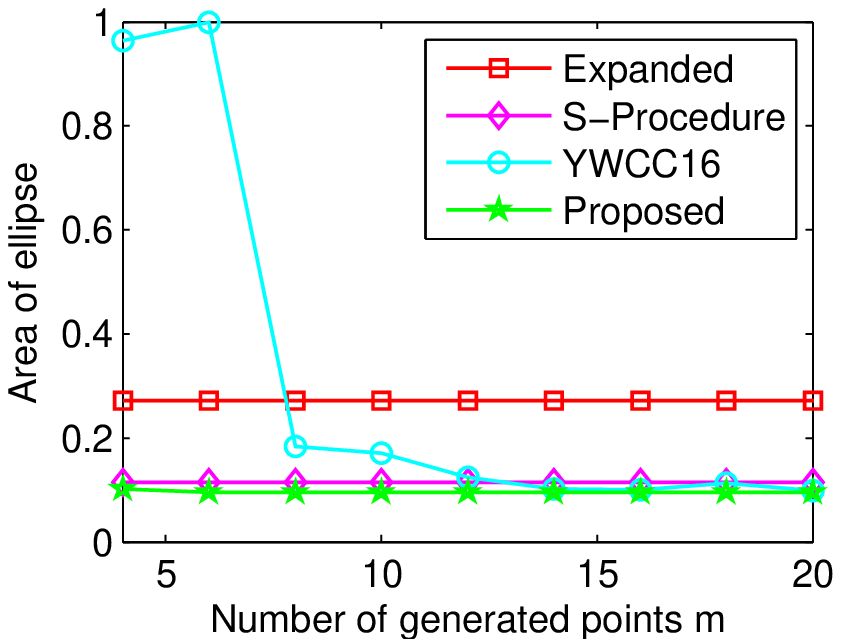}
}
\subfloat[]{
\includegraphics[width=55mm,height=45mm]{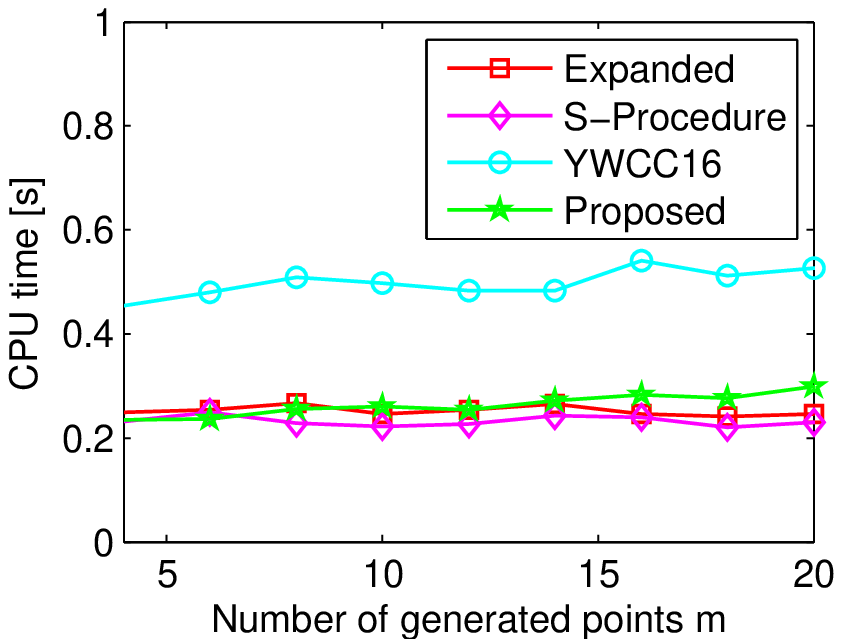}
}
\caption{Numerical test for M=5: (a) Areas of bounding ellipses, (b) CPU times.}
\label{fig:Ellipse_5}
\end{figure*}

In Fig. \ref{fig:Ellipse_5}-(a), we compare the area obtained by different methods for the scenario in which five ellipses intersect, i.e., $M=5$.
Once again, \emph{Proposed} outperforms all the other techniques.
\emph{YWCC16} faces degeneracy problems with the choice of $m=4$ and $m=6$, however, for the other values of $m$ it outperforms the \emph{S-Procedure} and \emph{Expanded}.
The computation times of different techniques are also in the same range except for \emph{YWCC16} which is higher, as shown in Fig. \ref{fig:Ellipse_5}-(b). Note that \emph{Proposed} obtains similar result to \emph{YWCC16}, with lower computational cost.

\begin{figure*}[h]
\centering
\subfloat[]{
\includegraphics[width=55mm,height=45mm]{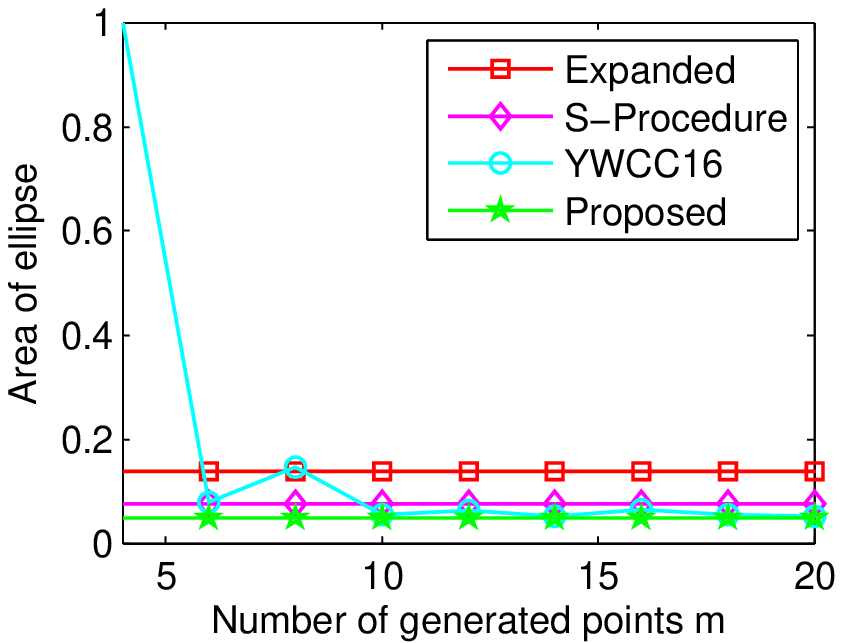}
}
\subfloat[]{
\includegraphics[width=55mm,height=45mm]{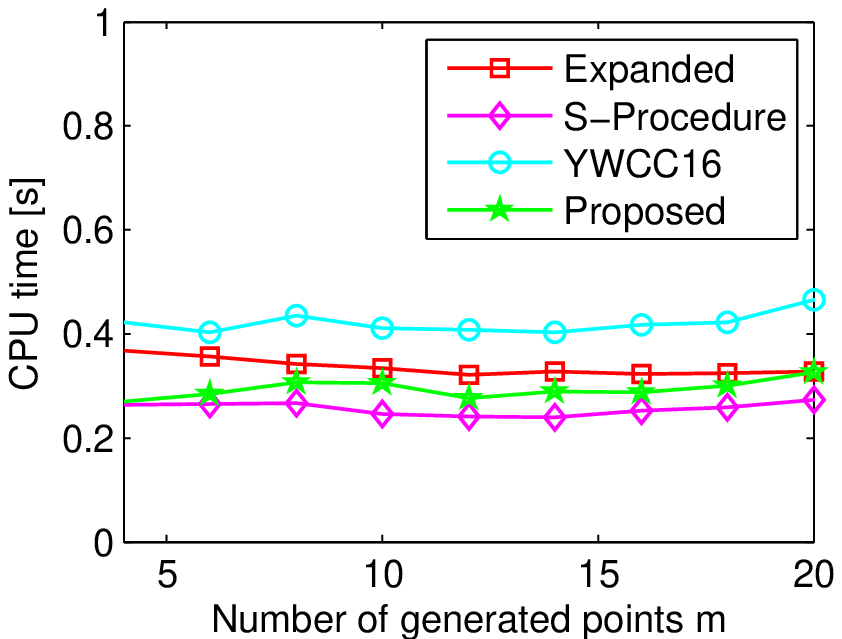}
}
\caption{Numerical test for M=10: (a) Areas of bounding ellipses, (b) CPU times.}
\label{fig:Ellipse_10}
\end{figure*}

In Fig. \ref{fig:Ellipse_10}-(a), the number of ellipses has increased to $M=10$.
Once again, \emph{Proposed} outperforms all the other techniques and \emph{S-Procedure} has very similar performance.
\emph{YWCC16} faces degeneracy problems with small $m$, however, for the large values of $m$, i.e., $m\geq10$ it yields a bounding ellipse, which has similar area compared to the one obtained by \emph{Proposed}.
The computation times of different techniques are also in the same range, as shown in Fig. \ref{fig:Ellipse_10}-(b), while \emph{S-Procedure} yields a slightly lower cost.
The computational cost of \emph{Proposed} is lower than \emph{YWCC16} when similar bounding ellipses are obtained, i.e., for $m \geq 10$.

\begin{figure*}[h]
\centering
\subfloat[]{
\includegraphics[width=55mm,height=45mm]{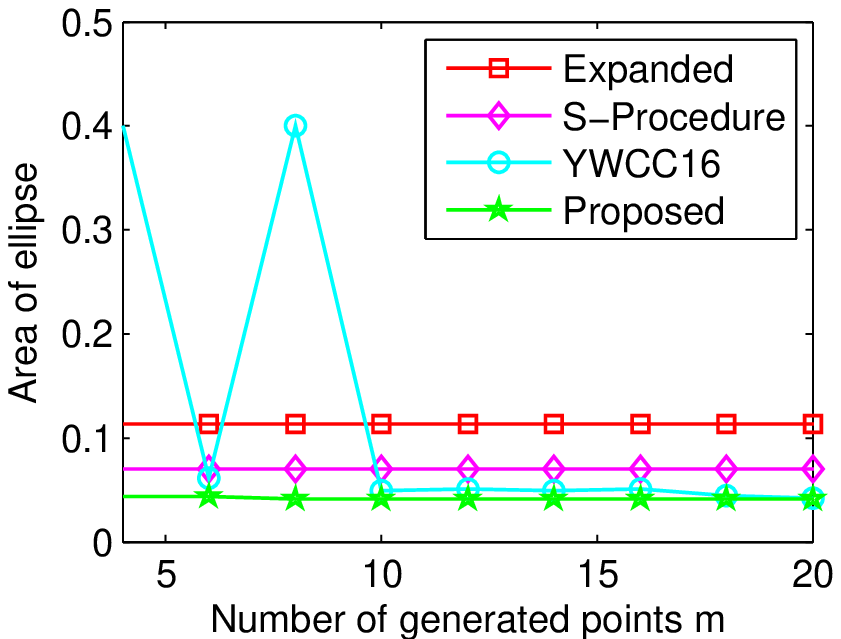}
}
\subfloat[]{
\includegraphics[width=55mm,height=45mm]{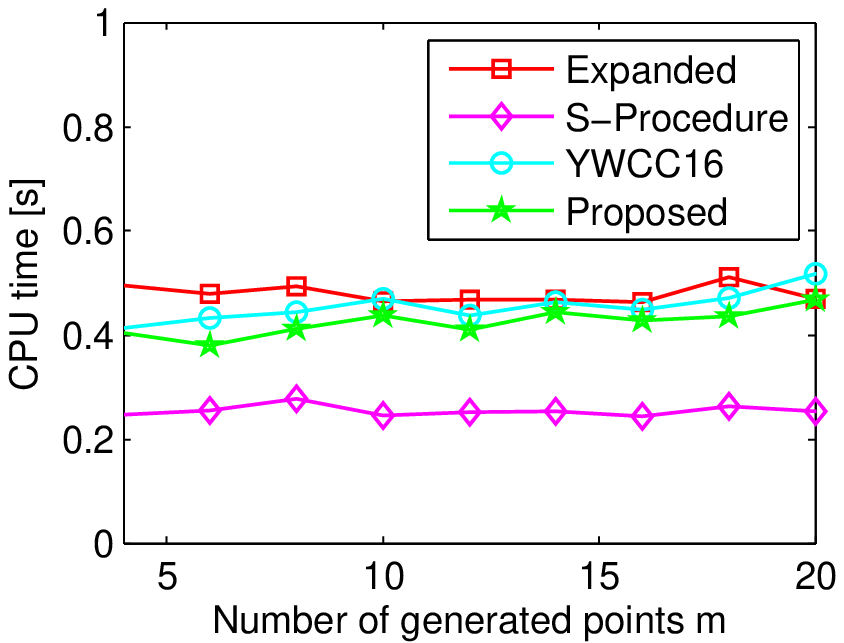}
}
\caption{Numerical test for M=20: (a) Areas of bounding ellipses, (b) CPU times.}
\label{fig:Ellipse_20}
\end{figure*}

In Fig. \ref{fig:Ellipse_20}-(a), the number of ellipses has increased to $M=20$.
Once again, \emph{Proposed} outperforms all the other techniques and \emph{YWCC16} has very similar performance when $m \geq 10$, while it faces degeneracy problems with some values of $m$ such as $m=4$ and $m=8$.
The computation times of different techniques are shown in Fig. \ref{fig:Ellipse_20}-(b).
The \emph{S-Procedure} yields the lowest cost compared to the others, while \emph{Proposed}, \emph{YWCC16} and \emph{Expanded} have higher computation time.
However, \emph{YWCC16} is not working properly for small value of $m$. Only for larger values of $m$ it starts yielding similar bounding ellipse.
When the number of ellipses grows, it takes ${\cal O}(M^3)$ for \emph{Proposed} to find the intersection points of ellipses.
Although this might seem a disadvantage of the proposed method, the application where the intersection of very large number of ellipses is required is very limited to the knowledge of the authors. 

\newpage

\emph{Remarks:}

\begin{itemize}

\item
From Fig. \ref{fig:Ellipse_2}-(a), Fig. \ref{fig:Ellipse_3}-(a), Fig. \ref{fig:Ellipse_5}-(a), Fig. \ref{fig:Ellipse_10}-(a), and Fig. \ref{fig:Ellipse_20}-(a), we observe that the area of the bounding ellipse obtained by \emph{YWCC16} does not always decrease when $m$ increases slightly. This is because some of the discrete points that are generated on ellipse $\xi_i$ that lie on the intersection region, may not necessarily be on the intersection region when $m$ is increased. This situation happens mostly when $m$ is increased by a small amount say 2 or 3.  
However, the area obtained by \emph{Proposed} does not increase because the number of discrete points on each arc segment of the feasible region remains either the same or increases when $m$ becomes larger.

\item
From Fig. \ref{fig:Ellipse_2}-(b), Fig. \ref{fig:Ellipse_3}-(b), Fig. \ref{fig:Ellipse_5}-(b), Fig. \ref{fig:Ellipse_10}-(b), and Fig. \ref{fig:Ellipse_20}-(b), we observe that the computation times of \emph{YWCC16} and \emph{Proposed} do not necessarily increase when $m$ is increased slightly.
The first reason is that sometimes when $m$ is increased slightly, the number of remaining discrete points on the intersection region does not necessarily increase and hence $m_p$ remains the same; thus the optimization problem in \eqref{Min_Vol_Ellipse_Points} does not change. 
Since the computation times of \emph{YWCC16} and \emph{Proposed} are also related to the number of constraints in \eqref{Min_Vol_Ellipse_Points} and this number might not change much by a slight increase in $m$, it is unlikely that the computation times increase noticeably.
%
The second reason is that the CVX optimization packages does not necessarily take the same amount of time to solve the same optimization problem.
%
This is because CVX uses an iterative interior point method that is initialized randomly and its convergence speed might be different every time it is utilized.
The computation times of \emph{S-Procedure} and \emph{Expanded} methods also show that although for different $m$ the same optimization problem is solved, the computation times are not exactly the same.

\end{itemize}

\begin{figure*}[htbp]
\centering
\subfloat[]{
\includegraphics[width=45mm,height=40mm]{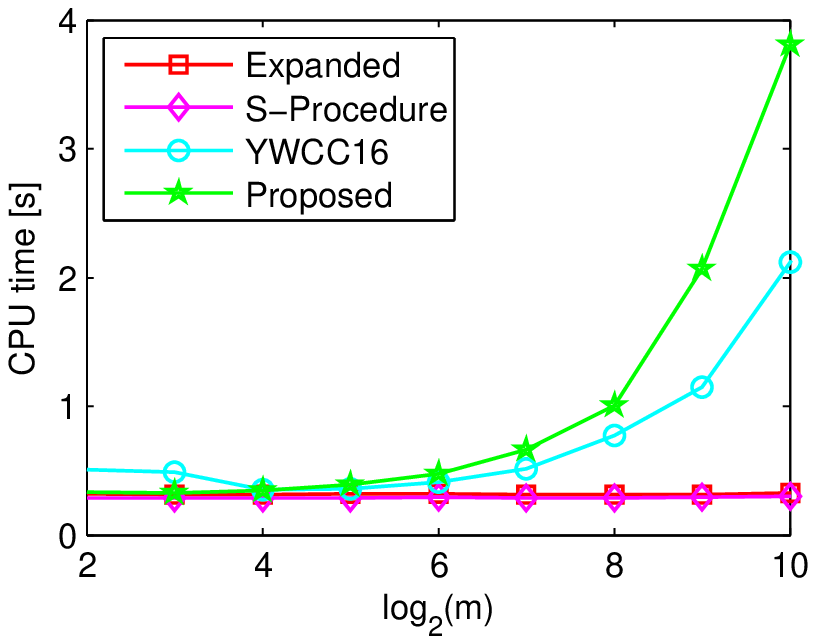}
}
~
\subfloat[]{
\includegraphics[width=45mm,height=40mm]{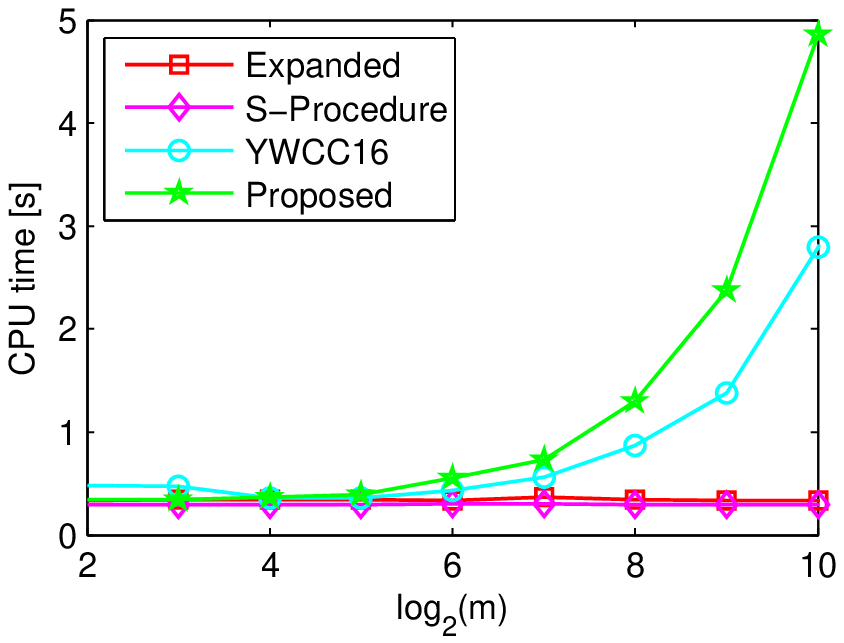}
}
\\
\subfloat[]{
\includegraphics[width=45mm,height=40mm]{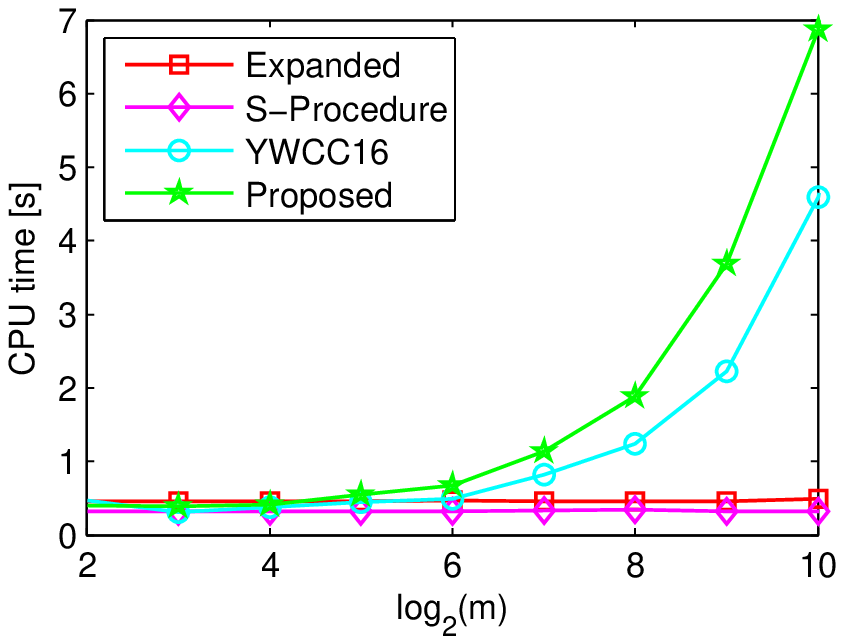}
}
~
\subfloat[]{
\includegraphics[width=45mm,height=40mm]{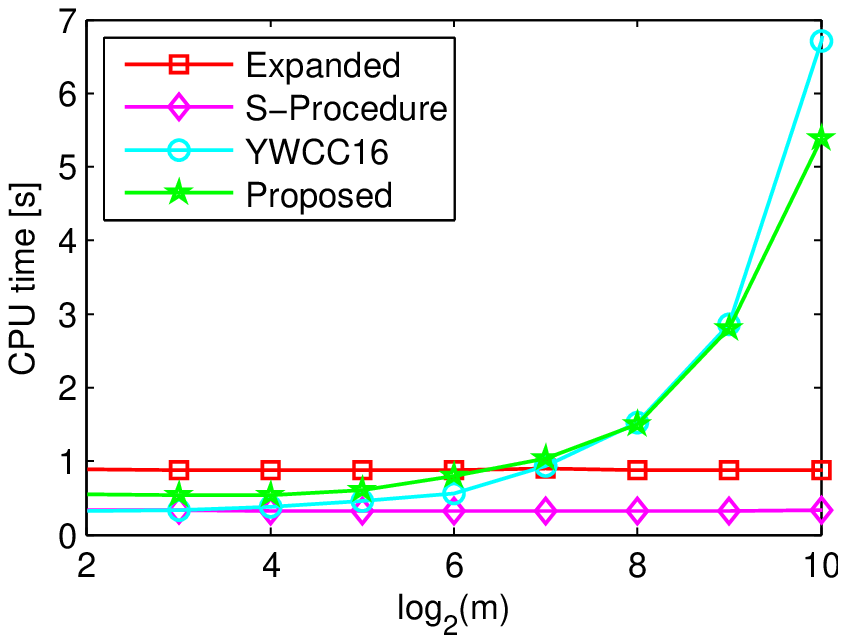}
}
\caption{Comparison of CPU time for different algorithms as a function of $ \mathrm{log}_2 (m) $ for the three scenarios with $M$ ellipses: (a) $M=3$, (b) $M=5$, (c) $M=10$, (d) $M=20$.}
\label{fig:CPU_Comparison}
\end{figure*}

Although the computation costs of \emph{Proposed} remained nearly constant for a small increase in $m$, we will show that when $m$ is increased by a larger amount, the computational cost grows noticeably.
The same behaviour is observed for \emph{YWCC16}.
In Fig. \ref{fig:CPU_Comparison}, the computation times of the different algorithms are plotted with respect to the logarithm of the number of discrete points, i.e., $\mathrm{log}_2(m)$.
Since the other benchmark approaches do not depend on $m$, their computational costs are almost fixed.
As observed, for small $m$, the computation times of \emph{YWCC16} and \emph{Proposed} are in the same range as those of \emph{Expanded-inner} and \emph{S-Procedure}.
However, by increasing $m$ further beyond 64, \emph{YWCC16} and \emph{YWCC16} will become more computationally demanding due to their dependence on the number of discrete points, both in the generation of the polygon, and in solving the optimization problem in \eqref{Min_Vol_Ellipse_Points}.
While this increase may seem to be a weak point of the proposed method, however, using large number of discrete points such as 1000 will be unnecessary for \emph{Proposed}.
This is because as the number of ellipses increases, the number of intersection points on $\cal E$ also increases and thus by only relying on these intersection points in order to find a polygon, a tight bounding ellipse can be obtained. Therefore, generating too many points on each arc segment between two neighbouring intersection points will not improve the tightness of the bounding ellipse noticeably.

\section{Conclusion} \label{Sec:Conclusion}
In this paper, we developed and studied tight outer approximation of the intersection region of a finite number of ellipses in 2-D space.
The main idea is to outer-approximate the feasible region by a tight polygon, and then find the smallest area ellipse containing the vertices of the polygon.
To find the polygon, we proposed to first find a set of discrete points on the boundary of the intersection region, and by linearizing the curves at those points find the half planes which form the polygon. 
In order to generate the discrete points on the boundary of the intersection region we first determined the intersection points, and then generated a required number of points on each segment of the intersection region connecting the two neighbouring points.
Through numerical experiments, it was illustrated that the proposed method could offer a tighter outer-approximation of the intersection of ellipses compared to the conventional methods found in the literature with similar computational cost.
Therefore, the proposed method, i.e., \emph{Proposed}, offers the best  trade-off between accuracy of the outer-approximation and computational cost, and hence will be preferred most of the time. 
In future work it may be worthwhile to program the techniques developed in \cite{Subexponential_bound} and \cite{Arrangements_curves} and apply them to the problem studied here to determine if they offer any practical advantages to the methods empirically investigated here.

\bibliographystyle{IEEEtran}

\bibliography{References}

\end{document}